\documentclass[aps,prc,twocolumn,tightenlines,superscriptaddress]{revtex4-1}
\usepackage{epsfig,rotating}
\usepackage{multirow}

\newcommand{\nuc}[2]{\hbox{$^{#1}$#2}}

\begin{document}
\title{Intermediate-energy inverse-kinematics one-proton pickup
  reactions on neutron-deficient $fp$-shell nuclei}
\author{S.\ McDaniel}
    \affiliation{National Superconducting Cyclotron Laboratory,
      Michigan State University, East Lansing, Michigan 48824}
    \affiliation{Department of Physics and Astronomy,
      Michigan State University, East Lansing, Michigan 48824}
\author{A.\ Gade}
   \affiliation{National Superconducting Cyclotron Laboratory,
      Michigan State University, East Lansing, Michigan 48824}
  \affiliation{Department of Physics and Astronomy,
     Michigan State University, East Lansing, Michigan 48824}
\author{J.\,A.\ Tostevin}
 \affiliation{Department of Physics, Faculty of Engineering and
      Physical Sciences, University of Surrey, Guildford,
      Surrey GU2 7XH, United Kingdom}
\author{T.\ Baugher}
    \affiliation{National Superconducting Cyclotron Laboratory,
      Michigan State University, East Lansing, Michigan 48824}
    \affiliation{Department of Physics and Astronomy,
      Michigan State University, East Lansing, Michigan 48824}
\author{D.\ Bazin}
    \affiliation{National Superconducting Cyclotron Laboratory,
      Michigan State University, East Lansing, Michigan 48824}
\author{B.\,A.\ Brown}
    \affiliation{National Superconducting Cyclotron Laboratory,
      Michigan State University, East Lansing, Michigan 48824}
    \affiliation{Department of Physics and Astronomy,
      Michigan State University, East Lansing, Michigan 48824}
\author{J.\ M.\ Cook}
    \affiliation{National Superconducting Cyclotron Laboratory,
      Michigan State University, East Lansing, Michigan 48824}
    \affiliation{Department of Physics and Astronomy,
      Michigan State University, East Lansing, Michigan 48824}
\author{T.\ Glasmacher}
    \affiliation{National Superconducting Cyclotron Laboratory,
      Michigan State University, East Lansing, Michigan 48824}
    \affiliation{Department of Physics and Astronomy,
      Michigan State University, East Lansing, Michigan 48824}
\author{G.\ F. \ Grinyer}\altaffiliation[Present address ]{GANIL,
CEA/DSM-CNRS/IN2P3, Bvd Henri Becquerel, 14076 Caen, France}
    \affiliation{National Superconducting Cyclotron Laboratory,
      Michigan State University, East Lansing, Michigan 48824}
\author{A.\ Ratkiewicz}
    \affiliation{National Superconducting Cyclotron Laboratory,
      Michigan State University, East Lansing, Michigan 48824}
    \affiliation{Department of Physics and Astronomy,
      Michigan State University, East Lansing, Michigan 48824}
\author{D.\ Weisshaar}
    \affiliation{National Superconducting Cyclotron Laboratory,
      Michigan State University, East Lansing, Michigan 48824}
\date{\today}

\begin{abstract}
{\bf Background:} Thick-target-induced nucleon-adding transfer reactions
onto energetic rare-isotope beams are an emerging spectroscopic tool.
Their sensitivity to single-particle structure complements one-nucleon
removal reaction capabilities in the quest to reveal the evolution of
nuclear shell structure in very exotic nuclei. {\bf Purpose:} To add
intermediate-energy, carbon-target-induced one-proton pickup reactions
to the arsenal of $\gamma$-ray tagged direct reactions applicable in
the regime of low beam intensities and to apply these for the first
time to $fp$-shell nuclei. {\bf Methods:} Inclusive and partial cross
sections were measured for the $\nuc{12}{C}(\nuc{48}{Cr},\nuc{49}{Mn}
+\gamma)$X and $\nuc{12}{C}(\nuc{50}{Fe},\nuc{51}{Co}+\gamma)$X proton
pickup reactions at 56.7 and 61.2~MeV/nucleon, respectively, using
coincident particle-$\gamma$ spectroscopy at the NSCL. The results are
compared to reaction theory calculations using $fp$-shell-model nuclear
structure input. For comparison with our previous work, the same reactions
were measured on \nuc{9}{Be} targets. {\bf Results:} The measured partial
cross sections confirm the specific population pattern predicted by
theory, with pickup into high-$\ell$ orbitals being strongly favored;
driven by linear and angular momentum matching. {\bf Conclusion:} Carbon
target-induced pickup reactions are well-suited, in the regime of modest
beam intensity, to study the evolution of nuclear structure, with specific
sensitivities that are well described by theory.
\end{abstract}

\maketitle
\section{Introduction}
For several decades, direct reactions have been crucial tools for the
investigation of the single-particle structures that underlie the
complex many-body problem of the atomic nucleus. At sufficiently high
collision energies one or two nucleons can be removed from the projectile
or transferred between the target and projectile nuclei in their fast,
surface-grazing collisions. Classic examples of such {\em spectroscopic}
direct reactions are light-ion-induced one- and two-nucleon transfer
reactions on thin, stable targets at incident energies of several MeV
per nucleon. Examples are the (d,p), (p,d), (p,t) and (t,p) reactions.
Here, it is the detection of the light charged particles in the exit
channel that allows the determination of final-state energies, cross
sections and their angular distributions, having sensitivity to the
location, spectroscopic strengths and the transferred orbital angular
momenta of the active one- and two-nucleon configurations (overlaps),
respectively.

With the advent of rare-isotope beam facilities, short-lived
neutron-rich or neutron-deficient nuclear species have become
available for nuclear structure studies. Many of the most exotic
nuclei can be produced efficiently by projectile fragmentation and
are available for experiments as fast heavy-ion beams with $v/c
\geq 0.30$. In this regime, of very exotic nuclei with modest beam
intensities of $1-10^3$ particles/second, luminosities comparable
to stable-beam experiments can be restored by employing thick
reaction targets and using in-beam $\gamma$-ray spectroscopy to
tag the final states of the projectile-like reaction residues
\cite{Gad08a}. Over the past decade, $\gamma$-ray tagged, thick
\nuc{9}{Be} target-induced one- and two-nucleon knockout reactions
in inverse kinematics, such as \nuc{9}{Be}(\nuc{A}{Z},\nuc{A-1}
{Z}$+\gamma$)X and \nuc{9}{Be}(\nuc{A}{Z},\nuc{A-2}{(Z$-$2)}$+
\gamma$)X, have been developed as spectroscopic reaction tools.
In the low intensity, fast secondary beam regime these reaction
mechanisms can probe in some detail the single-particle content
of the nuclear many-body wave function at and near the Fermi
surface(s) \cite{Han03}.

In these reactions, one-nucleon removal selectively populates
hole-like configurations with respect to the projectile ground
state. Similar to transfer reactions, they probe the location of
such single-hole states, their spectroscopic strengths and orbital
angular momenta. Sudden two-proton (neutron) removal reactions from
neutron (proton)-rich nuclei also proceed as direct reactions. They
probe, specifically, the parentage and phase of the participating
two-nucleon configurations near the well-bound Fermi surface of the
ground state of the projectile, built upon the final states of the
projectile-like reaction residues~\cite{Baz03,Tos0406,Yon06,Sim09}.
Details of this nuclear structure model information is carried by
the two-nucleon amplitudes (overlaps), see e.g. \cite{Tos0406}.

To complement this arsenal of fast-beam, inverse kinematics nucleon
removal reactions, one-nucleon pickup reactions onto fast projectile
beams from thick carbon and beryllium targets have recently been
studied as a spectroscopic probe sensitive to {\it particle}-like
states/configurations \cite{Gade07a,Gade07b,Gade11}. The choice of
reaction target is now motivated by generic linear and angular momentum
matching considerations and that, at high collision energies, the
pickup of well-bound target nucleons is favored \cite{Shi05,Mic06}.
In common with the thick-target nucleon removal experiments, the
energy resolution of the populated final-states relies on $\gamma$-ray
spectroscopy performed in coincidence with the detection of the heavy
projectile-like pickup reaction residues. However, unlike in removal
reactions, and rooted in the two-body nature of the reaction mechanism,
the longitudinal momentum distribution of the projectile-like reaction
residues carries no useful information on the orbital angular momentum
of the transferred nucleon. Rather, it can be used to estimate whether
there are significant non-two-body final-state contributions to the
measured reaction yields~\cite{Gade11}.

So far, thick \nuc{12}{C} and \nuc{9}{Be} targets were employed and
compared \cite{Gade11} for the fast one-neutron pickup studies. Only
\nuc{9}{Be} targets have been used to date for measured one-proton
pickup cases \cite{Gade07a,Gade07b}. The goals of the present work
are two-fold. We investigate:\smallskip \\
\noindent (i)
the use of a \nuc{12}{C} target to induce the fast-beam one-proton
pickup reaction of type $\nuc{12}{C}(\nuc{A}{Z},\nuc{A+1}{(Z+1)}+
\gamma)$, including the measurement and theoretical description of
partial cross sections, and\smallskip \\
\noindent (ii) extension of the study of this spectroscopic tool
into $fp$-shell systems.

An important aspect of the use of pickup onto fast projectile beams
is the anticipated selectivity of this transfer process to high-$\ell$
orbitals, or fragments thereof, in the projectile residues, promising
this technique as suited for identifying intruder states that are
direct indicators of shell evolution. This makes projectiles with
vacancies in the proton $1f_{7/2}$ and $2p_{3/2,1/2}$ orbits a
logical choice to advance such one-proton pickup developments, to
highlight and contrast the pickup yields into these $\ell$=3 and
$\ell$=1 single-proton configurations.

For the present study we use a cocktail beam of projectiles containing
the $N$=24 isotones \nuc{48}{Cr} and \nuc{50}{Fe}. We will focus on the
\nuc{12}{C}(\nuc{48}{Cr},\nuc{49}{Mn}+$\gamma$)X and \nuc{12}{C}(\nuc{50
}{Fe},\nuc{51}{Co}+$\gamma$)X reactions at mid-target energies of 56.7
and 61.2 MeV/nucleon, respectively. To allow comparison with the earlier
work (that used \nuc{9}{Be} targets) the \nuc{9}{Be}(\nuc{48}{Cr},\nuc{49
}{Mn}+$\gamma$)X and \nuc{9}{Be}(\nuc{50}{Fe},\nuc{51}{Co}+$\gamma$)X
reactions, at mid-target energies of 50.7 and 54.8 MeV/nucleon,
respectively, were also measured with the same experimental setup.

\section{Experiments}
The projectile beam containing \nuc{48}{Cr} and
\nuc{50}{Fe} was produced by fragmentation of
a 160-MeV per nucleon \nuc{58}{Ni} primary beam provided by the
Coupled Cyclotron Facility at the National Superconducting Cyclotron
Laboratory (NSCL) at Michigan State University. The 893~mg/cm$^2$
thick \nuc{9}{Be} production target was
located at the mid-acceptance target position of NSCL's A1900 fragment
separator~\cite{a1900}. An achromatic Al wedge of
300-mg/cm$^2$ thickness together with slit systems were used to
optimize the beam purity on \nuc{50}{Fe}. The resulting cocktail beam
of $N=24$ isotones was
transmitted to the reaction target position of the S800 spectrograph at
typical rates of 2500, 2000, and
500 particles per second and purities of 39\%, 40\%, and 11\%,
respectively, for \nuc{48}{Cr}, \nuc{49}{Mn}, and \nuc{50}{Fe} at
$\Delta p/p=1$\% total A1900 momentum acceptance.

Reaction targets of 72.8(13)-mg/cm$^2$ thick
\nuc{12}{C} and 188(4)-mg/cm$^2$ thick \nuc{9}{Be} were used at the
target position of the S800 spectrograph~\cite{s800} for the different
settings. The
high-resolution $\gamma$-ray detection system SeGA~\cite{sega}, an
array of
32-fold segmented high-purity germanium detectors, was used to detect
the de-excitation $\gamma$ rays emitted by the reaction
residues. Sixteen of the SeGA detectors were arranged in two rings with
relative angles of 90$^{\circ}$ (9 detectors) and 37$^{\circ}$ (7
detectors) with respect to the beam axis. The
high degree of segmentation
allows for event-by-event Doppler reconstruction of the $\gamma$
rays emitted by the projectile-like reaction residues in flight. The
angle of the $\gamma$-ray emission that enters the Doppler
reconstruction is deduced from the position of the segment that
registered the largest $\gamma$-ray energy deposition. The photopeak
efficiency of the detector array was calibrated with standard sources and
corrected for the Lorentz boost of the $\gamma$-ray
distribution emitted by the projectile-like reaction residues in
flight at $v/c>0.25$.

\begin{figure}[h]
\epsfxsize 7.2cm
\epsfbox{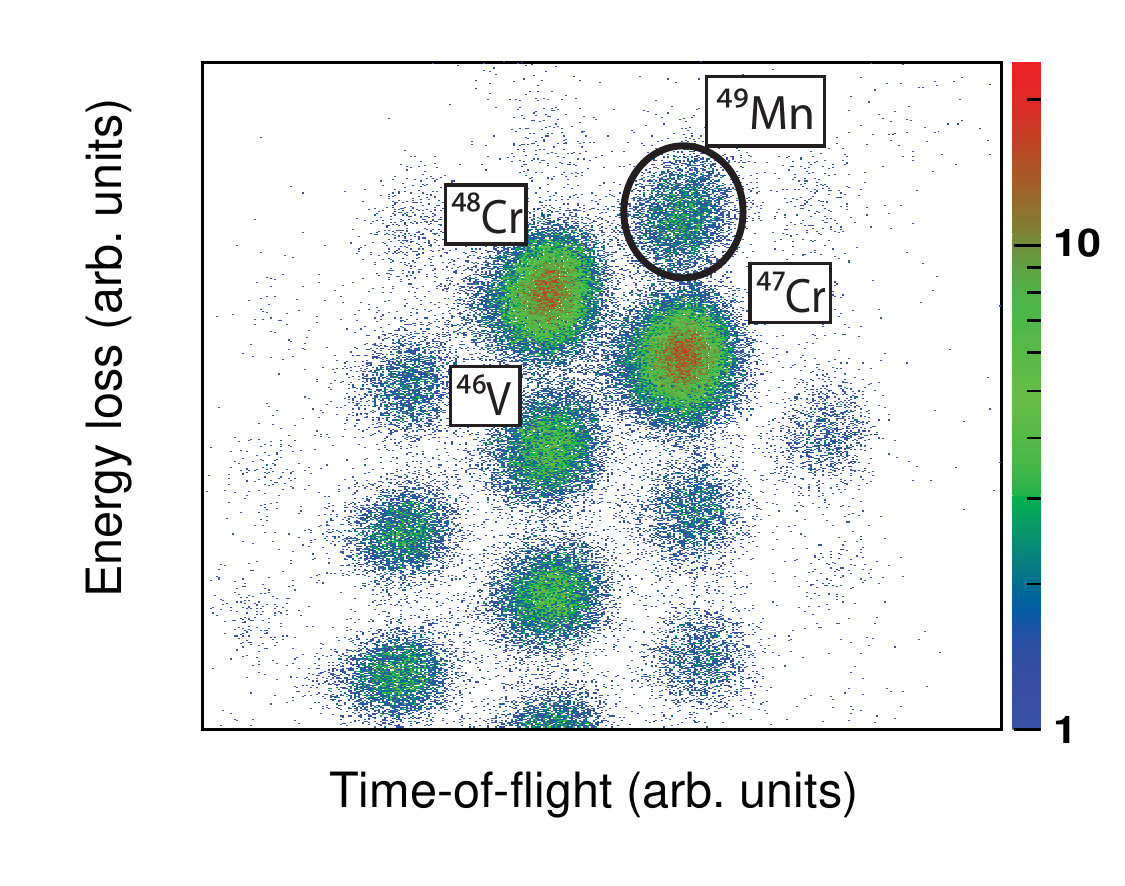}
\caption{\label{fig:pid} (Color online) Particle identification for
the reaction residues in the \nuc{9}{Be}(\nuc{48}{Cr},\nuc{49}{Mn})X
setting. Plotted is the energy loss measured with the S800
ionization chamber versus the time of flight measured between the
plastic trigger scintillator at the back of the S800 focal plane and
a scintillator at the spectrograph's object location.}
\end{figure}

Reaction residues were identified event-by-event with the
focal-plane detection system~\cite{Yurkon99} of the S800
spectrograph. The energy loss measured in the S800 ionization
chamber and flight times, corrected for the angle and momentum of each nucleus,
were used to identify the projectile-like reaction
residues exiting the target. The incoming projectiles were
identified from their difference in flight time measured with two plastic
scintillators upstream of the target. The particle-identification spectrum for
\nuc{49}{Mn} produced in \nuc{48}{Cr} + \nuc{12}{C} is shown in
Fig.~\ref{fig:pid}. The projectile beams passing through the target
are close in magnetic rigidity to the respective one-proton pickup residues and
the beam blocker located at the entrance of the S800 focal plane had
to be used to block a large fraction of the unreacted
projectiles. Still, the most intense contribution in the
identification spectrum in Fig.~\ref{fig:pid} is from unreacted
\nuc{48}{Cr}.

For each target and projectile beam, the inclusive cross sections for
the one-proton pickup reaction to all projectile + proton bound
final states were deduced from
the yield of detected pickup residues divided by the number of
incoming projectiles relative to the number density of the
\nuc{9}{Be} and \nuc{12}{C} reaction targets,
respectively. The deduced inclusive cross sections are
$\sigma_{exp}^{inc}(\nuc{49}{Mn})=2.00(13)$~mb and
$\sigma_{exp}^{inc}(\nuc{51}{Co})=0.53(13)$~mb for the
\nuc{12}{C}-induced pickups and
$\sigma_{exp}^{inc}(\nuc{49}{Mn})=1.63(8)$~mb and
$\sigma_{exp}^{inc}(\nuc{51}{Co})=0.57(8)$~mb for the
\nuc{9}{Be}-induced pickups. The quoted uncertainties combine
statistical and systematic
errors, including the stability of the secondary beam composition, the
choice of the software gates, and corrections for the acceptance
losses in the tails of the residue parallel momentum distributions.

Event-by-event in-beam $\gamma$-ray spectroscopy in coincidence
with detection of the projectile-like pickup residue allowed the
identification of excited final states and measurement of their
population in the pickup to \nuc{49}{Mn}. No bound excited states were
observed for \nuc{51}{Co}, for which a proton separation energy of
only $S_p \approx 90$~keV is predicted from systematics~\cite{AW2003}.

Partial cross sections to individual final
states were obtained from the efficiency-corrected full-energy peak
areas in the Doppler reconstructed $\gamma$-ray spectra relative to
the number of pickup products, and were corrected for
feeding. Unfortunately, the statistics were not sufficient to tag
the final state of the \nuc{11}{B} and \nuc{8}{Li} target residues
in the laboratory-frame $\gamma$-ray spectra (see~\cite{Gade07a} for
a higher statistics $sd$-shell case for which it was possible to
tag the \nuc{8}{Li} target residues left in the first excited state).

\section{Theoretical reaction model\label{reaction}}
The one-proton transfers, from \nuc{12}{C} and \nuc{9}{Be}, are assumed
to take place in a single step from the ground states of these targets
to bound proton single-particle states built on the ground states of
the projectiles. So, as was used in earlier work, the one-proton pickup
single-particle cross sections are calculated using the post-form of
the distorted waves Born approximation (DWBA) to the pickup reaction
transition amplitude. The transfer interaction (that binds the transferred
proton and the target residues \nuc{11}{B} and \nuc{8}{Li}) is treated
in finite-range.  The calculations assume that both the entrance and
exit channels are two-body and hence that the proton is transferred
to bound states in the projectile-like residue and the target-like
residues remain bound. All calculations were performed using the
direct reactions code {\sc fresco} \cite{fresco}.

The single-particle cross sections (computed for a proton + projectile
spectroscopic factor of one) and the shell-model spectroscopic factors
enter the calculation of the theoretical (angle integrated) pickup
cross sections that are compared to the experimental results. For each
final state, $J^{\pi}$, of the projectile-like residue the theoretical
cross section is
\begin{equation}
\label{eq:xsec}
\sigma_{th}(J^{\pi})=\left(\frac{A_p}{A_p+1}\right)^{\!\!{\cal N}}\,
S_{SM}(J^{\pi})\,\sigma_{sp}(J^{\pi})\ .
\end{equation}
The first factor, that depends on the projectile mass $A_p$ and the
principal harmonic oscillator quantum number, ${\cal N}$, of the transferred
proton orbit, is a small (near unity) center-of-mass correction to the
shell-model spectroscopic factors, $S_{SM}$. ${\cal N}$=3 for the present
proton $fp$-shell cases. The measured cross sections and the calculated
(single-particle) cross sections $\sigma_{sp}(J^{\pi})$ to specific
final states in \nuc{49}{Mn} and \nuc{51}{Co} are both inclusive with
respect to the final states of the target residues. The calculations
thus include contributions from transitions involving several \nuc{11}{B}
and \nuc{8}{Li} final states. These will be detailed in the following
subsections for the \nuc{12}{C} and \nuc{9}{Be} target cases.

The spectroscopic amplitudes of the proton + projectile bound state
overlaps, $[\nuc{48}{Cr}(0^+)\otimes n\ell_j]$ and $[\nuc{50}{Fe}(0^+)
\otimes n\ell_j]$, were taken from $fp$-shell-model calculations using
the GXPF1A effective interaction~\cite{Honma}. The associated proton + projectile bound
state wave functions (radial overlaps) were calculated in real Woods-Saxon
potential wells with diffuseness parameter $a_0=0.7$ fm and spin-orbit
interaction strength of 6 MeV. The reduced radius parameters, $r_0$, of
these potentials were adjusted to reproduce the root mean squared (rms)
radii of each proton single-particle orbital given by spherical Hartree
Fock calculations, as has been discussed in detail elsewhere \cite{Gad08b}.
These same Hartree-Fock (HF) calculations were used to calculate the
projectile and projectile-residue densities, for calculation of the
channel distorting potentials, as will be discussed below. The $r_0$
values were 1.24 and 1.25 fm for the $1f_{7/2}$ and $2p$-shell proton
orbitals in \nuc{49}{Mn}, respectively, and 1.23 fm for the $1f_{7/2}$
proton orbital in \nuc{51}{Co}.

\subsection{Reactions on \nuc{12}{C} target}
The proton is assumed to be transferred in a single step from a bound
state in the \nuc{12}{C}(g.s.) into a proton + projectile bound state.
The \nuc{11}{B} target residue is assumed to be left in one of three
bound states, the $3/2^-_1$ ground state, the $1/2^-_1$ excited state
at 2525~keV, and the $3/2^-_2$ excited state at 5020~keV. The shell-model
spectroscopic factors (WBP effective interaction \cite{wbp}) for these
three states essentially exhaust the expected spectroscopic strength,
of 4, being $3.16$, 0.58, and 0.19, respectively. The residual strength,
of 0.07, is distributed over many states up to the proton separation
energy of \nuc{11}{B} and is neglected. Woods-Saxon potentials are used
to generate these proton + $\nuc{11}{B}$ bound states, their geometry
taken from Ref.\ \cite{Brown02}. These have reduced radius parameter
$r_0=1.310$ fm and diffuseness $a=0.55$ fm.

The nuclear distorting interactions in the \nuc{12}{C} + projectile
(entrance) and \nuc{11}{B} + pickup residue (exit) channels were
calculated. We adopted the method used in earlier fast nucleon removal
reaction studies \cite{Gad08b}, i.e. by double folding the point neutron
and proton densities of the projectiles and residues, \nuc{48}{Cr},
\nuc{49}{Mn} and \nuc{50}{Fe}, \nuc{51}{Co}, and of \nuc{12}{C}, \nuc{11}{B}
with an effective nucleon-nucleon ($NN$) interaction~\cite{Tos0406}. For
the projectile-like systems the densities were obtained from spherical
Skyrme (SkX interaction) Hartree-Fock (HF) calculations \cite{skx}. The
target-like systems, \nuc{12}{C} and \nuc{11}{B}, were assumed to have
Gaussian density distributions with rms radii of 2.32 fm and 2.11 fm.

The proton pickup reactions to \nuc{49}{Mn} were thus computed as
$\nuc{48}{Cr}(\nuc{12}{C},\nuc{11}{B}(I^{\pi}))\nuc{49}{Mn}(J^{\pi})$
leading to the $I^{\pi}=3/2^-_{1}$, $1/2^-_1$, and $3/2^-_2$, \nuc{11}{B}
final states. The first excited $J^{\pi}=7/2^-$, $1/2^-$, and $3/2^-$
final states of \nuc{49}{Mn} were considered.  Pickup to the $5/2^-$
ground state of \nuc{49}{Mn} is negligible due to the very small
shell-model spectroscopic factor, $S_{SM}=0.002$. For the (as yet
unobserved) $1/2^-$ and $3/2^-$ excited states we used the level
energies of the corresponding states in the \nuc{49}{Cr} mirror
system. The energy of the $7/2^-_1$ state is known to be 262~keV.

Table~ \ref{tab:xsec_c_49} summarizes the inputs to the reaction
calculations. There, $S_p^{\mathrm{r.eff}}=S_p(\nuc{49}{Mn}) - E_x
(\nuc{49}{Mn})$ is the effective proton separation energy from the
projectile-like residue, $S_p^{\mathrm{t.eff}}=S_p(\nuc{12}{C})+E_x
(\nuc{11}{B})$ that from the target. The single-particle proton pickup
cross section to each final state of \nuc{49}{Mn} is then the sum of
the cross sections to the three \nuc{11}{B} final states. As
expected, based on the WBP shell-model spectroscopic factors shown
above, the \nuc{11}{B} ground state transition dominates in these
individual single-particle cross sections, $\sigma_{sp}$. It is
already clear from Table~\ref{tab:xsec_c_49} that our high projectile
beam energy strongly favors $\ell=3$ pickup, to the $1f_{7/2}$ proton
state of \nuc{49}{Mn}, over $\ell=1$ pickup to the other final states.
This is understood intuitively by generic, semi-classical linear and
angular momentum (mis)matching arguments, e.g. \cite{Brink72,Phil77}.

\begin{table}
\caption{Inputs used in the reaction model calculations for
\nuc{12}{C}(\nuc{48}{Cr},\nuc{49}{Mn})\nuc{11}{B}($I^{\pi}$). Shown
are the \nuc{49}{Mn} final-state angular momenta and parities, $J^{\pi}$,
their excitation energies, $E_x$, proton configurations, $n \ell_j$, and
effective proton separation energies, $S^{\mathrm{r.eff}}_p$. The
\nuc{11}{B} final-state spins and parities, $I^{\pi}$, effective proton
separation energies, $S_p^{\mathrm{t.eff}}$, and the single-particle
cross sections, $\sigma_{sp}$, for each transition are also shown. }
\begin{ruledtabular}
\begin{tabular}{lccccccc}
\multicolumn{4}{c}{\nuc{49}{Mn}}&\multicolumn{2}{c}{Target
  $\nuc{12}{C} \rightarrow \nuc{11}{B}$} & \\
$J^{\pi}$  & $E_x$ & $n\ell_j$ & $S_p^{\mathrm{r.eff}}$ & $I^{\pi}$ &
  $S_p^{\mathrm{t.eff}}$  &  $\sigma_{sp}$ \\
          & (keV) &    & (MeV) &   & (MeV) & (mb) \\
\hline
   $7/2^-$ & 262 & $1f_{7/2}$ & 1.823 & $3/2^-$ & 15.957 & 3.724 \\
     &     &           &        & $1/2^-$ & 18.082 &  0.238  \\
     &     &             &        & $3/2^-$ & 20.977 &   0.179  \\

$1/2^-$& 1703 & $2p_{1/2}$ & 0.382 & $3/2^-$ & 15.957 & 0.029\\
       &     &            &       & $1/2^-$ & 18.082   &0.010  \\
     &     &     &           & $3/2^-$ & 20.977   & 0.001 \\
   $3/2^-$ & 1741 & $2p_{3/2}$ & 0.344 & $3/2^-$ & 15.957 & 0.103  \\
     &     &     &               & $1/2^-$ & 18.082   & 0.008  \\
     &     &     &                & $3/2^-$ & 20.977   & 0.003
\label{tab:xsec_c_49}
\end{tabular}
\end{ruledtabular}
\end{table}

Pickup to the weakly-bound dripline nucleus \nuc{51}{Co} was computed
similarly, as $\nuc{50}{Fe}(\nuc{12}{C},\nuc{11}{B}(I^{\pi}))
\nuc{51}{Co}$, leading to the same \nuc{11}{B} final states. Only pickup
to the \nuc{51}{Co}($7/2^-$) ground state was considered due to the low
proton separation energy of $S_p(\nuc{51}{Co})=90$ keV \cite{AW2003}.
The calculations of the single-particle cross sections are summarized
in Table \ref{tab:xsec_c_51}.
\begin{table}
\caption{As Table~\ref{tab:xsec_c_49} but for \nuc{12}{C}(\nuc{
50}{Fe},\nuc{51}{Co})\nuc{11}{B}($I^{\pi}$).}
\begin{ruledtabular}
\begin{tabular}{lcccccccc}
\multicolumn{4}{c}{\nuc{51}{Co}}&\multicolumn{2}{c}{Target
  $\nuc{12}{C} \rightarrow \nuc{11}{B}$} &\\
$J^{\pi}$  & $E_x$ & $n\ell_j$ & $S_p^{\mathrm{r.eff}}$ & $I^{\pi}$ &
  $S_p^{\mathrm{t.eff}}$  &  $\sigma_{sp}$ \\
          & (keV) &    & (MeV) &   & (MeV) & (mb) \\
\hline
$7/2^-$ & $0  $ & $1f_{7/2}$ & 0.090 & $3/2^-$ & 15.957 &  2.708 \\
        &     &           &        & $1/2^-$ &  18.082   &0.165  \\
        &     &             &        & $3/2^-$ & 20.977   & 0.129
\label{tab:xsec_c_51}
\end{tabular}
\end{ruledtabular}
\end{table}

These calculated partial and summed theoretical cross sections, including
the shell model spectroscopic factors and the center-of-mass corrections,
are summarized in Tables~\ref{tab:conf_49} and~\ref{tab:conf_51}. These
results will be discussed and compared to the experimental data in Section
\ref{discussion}.

\subsection{Reactions on \nuc{9}{Be}}
The theoretical description of the \nuc{9}{Be}-target-induced one-proton
pickup reactions follows our earlier implementation, as was presented
in~\cite{Gade07a}. In this instance two bound and one unbound final state
are expected to be strongly populated in the \nuc{8}{Li} target residue
and will be considered. These are the $2^+$ ground state, the $1^+_1$
excited state at 980~keV, and the $3^+_1$ state at 2255(3)~keV, the
latter just above the first neutron threshold of $S_n(\nuc{8}{Li})=
2032.62(12)$~keV \cite{AW2003}. The $^{9}$Be(${3/2}^-)$ to $^8$Li$(I^+)$
radial overlaps and their associated spectroscopic amplitudes used
the Variational Monte Carlo (VMC) calculations of Wiringa {\em et al.}
\cite{BobW}, as were discussed in Ref.\ \cite{Gade07a}.

As for the \nuc{12}{C} target, the nuclear distorting interactions in
the entrance and exit channels were obtained by double folding the point
neutron and proton HF densities of the projectiles and their residues
and of \nuc{9}{Be} and \nuc{8}{Li} with an effective $NN$ interaction
\cite{Tos0406}. Both \nuc{9}{Be} and \nuc{8}{Li} were assumed to have
Gaussian density distributions with an rms radius of 2.36 fm.

The pickup reaction to \nuc{49}{Mn} was computed as $\nuc{48}{Cr}(
\nuc{9}{Be},\nuc{8}{Li}(I^{\pi}))\nuc{49}{Mn}(J^{\pi})$ to the three
final states of \nuc{8}{Li} outlined above. The excited $J^{\pi}=
7/2^-_1$, $1/2^-_1$ and $3/2^-_1$ \nuc{49}{Mn} final states were
considered. The input parameters and calculated cross sections are
summarized in Table \ref{tab:xsec_be_49}, with the same entries as
explained in the previous subsection. In this case the spectroscopic
strengths of the different \nuc{8}{Li} final states (for values, see
\cite{Gade07a}) are more equal and the ground state transition is
less dominant. As previously, the one-proton pickup cross section
to a particular \nuc{49}{Mn} final state is the sum of these
single-particle cross sections for the three \nuc{8}{Li} final
states. The calculations confirm the selectivity of pickup to
the high-$\ell$ final-state configuration.

\begin{table}
\caption{As Table~\ref{tab:xsec_c_49} but for \nuc{9}{Be}(
\nuc{48}{Cr},\nuc{49}{Mn})\nuc{8}{Li}($I^{\pi}$).}
\begin{ruledtabular}
\begin{tabular}{lccccccc}
\multicolumn{4}{c}{\nuc{49}{Mn}}&\multicolumn{2}{c}{Target
  $\nuc{9}{Be} \rightarrow \nuc{8}{Li}$} &\\
$J^{\pi}$  & $E_x$ & $n\ell_j$ & $S_p^{\mathrm{r.eff}}$ & $I^{\pi}$ &
  $S_p^{\mathrm{t.eff}}$  &   $\sigma_{sp}$ \\
          & (keV) &    & (MeV) &  & (MeV) & (mb)\\
\hline
   $7/2^-$ & 262 & $1f_{7/2}$ & 1.823 & $2^+$ & 16.888 &  2.213 \\
     &     &           &        & $1^+$ & 17.870 & 0.882  \\
     &     &             &        & $3^+$ & 19.143 &  0.811  \\
$1/2^-$& 1703 & $2p_{1/2}$ & 0.382 & $2^+$ & 16.888 &  0.021\\
       &     &            &       & $1^+$ & 17.870 & 0.014  \\
     &     &     &           & $3^+$ & 19.143 &  0.004\\
   $3/2^-$ & 1741 & $2p_{3/2}$ & 0.344 & $2^+$ & 16.888 &  0.049 \\
     &     &     &               & $1^+$ & 17.870 &  0.020 \\
     &     &     &                & $3^+$ & 19.143 &  0.014
\label{tab:xsec_be_49}
\end{tabular}
\end{ruledtabular}
\end{table}

The proton pickup to \nuc{51}{Co}, computed as $\nuc{50}{Fe}(\nuc{9}{Be},
\nuc{8}{Li}(I^{\pi}))\nuc{51}{Co}(7/2^-_{gs})$ was treated similarly. The
details of the calculations are summarized in Table~\ref{tab:xsec_be_51}.

\begin{table}
\caption{As Table~\ref{tab:xsec_c_49} but for \nuc{9}{Be}(\nuc{50}{Fe},
\nuc{51}{Co})\nuc{8}{Li}($I^{\pi}$).}
\begin{ruledtabular}
\begin{tabular}{lccccccc}
\multicolumn{4}{c}{\nuc{51}{Co}}&\multicolumn{2}{c}{Target
  $\nuc{9}{Be} \rightarrow \nuc{8}{Li}$} & \\
$J^{\pi}$  & $E_x$ & $n\ell_j$ & $S_p^{\mathrm{r.eff}}$ & $I^{\pi}$ &
  $S_p^{\mathrm{t.eff}}$  &   $\sigma_{sp}$ \\
          & (keV) &    & (MeV) & &  (MeV) & (mb) \\
\hline
$7/2^-$ & $0$ & $1f_{7/2}$ & 0.090 & $2^+$ & 16.888  & 1.625 \\
        &     &           &          & $1^+$ & 17.870  & 0.640 \\
        &     &           &          & $3^+$ & 19.143  & 0.593
\label{tab:xsec_be_51}
\end{tabular}
\end{ruledtabular}
\end{table}

\section{Results and Discussion\label{discussion}}
We now discuss the experimental results from the reactions induced
on the even-even \nuc{48}{Cr} and \nuc{50}{Fe} projectiles in comparison
to the calculations presented above. We also analyze the longitudinal
momentum distributions of the \nuc{49}{Mn} and \nuc{51}{Co} pickup
residues from the reactions on the \nuc{12}{C} target with emphasis
on confirming the (assumed) two-body nature of the reaction mechanism.

\subsection{Proton pickup to \nuc{49}{Mn}}
Neutron deficient \nuc{49}{Mn} is well-suited to further benchmark
the spectroscopic value of fast-beam one-proton pickup reactions.
Having six neutrons less than the only stable Mn isotope, its low
proton separation energy of $S_p=2085(25)$~keV~\cite{AW2003} limits
the number of proton-bound states and so minimizes the impact of
possible (unobserved) indirect feeding on the determination of the
partial cross sections from $\gamma$-ray spectroscopy. Furthermore,
the proton shell structure of \nuc{49}{Mn} allows one to examine
the high-$\ell$ orbital selectivity of the pickup reaction,
predicted to be strongly favored by momentum matching at high
projectile energy. The yrast level scheme of \nuc{49}{Mn}, discovered
in 1970 \cite{Cerny70}, is known up to high spin values from extensive
studies in the 90's focused on isospin symmetry \cite{OLea97,Cam90}.
Two low-lying states, dominated by single-particle configurations
involving the proton $2p_{1/2}$ and $2p_{3/2}$ orbits, are expected
around 1700 and 1740~keV, respectively, from comparison to the mirror
nucleus \nuc{49}{Cr}~\cite{nndc}, but have not yet been observed.
The most recent work on \nuc{49}{Mn} employed $\beta$-delayed proton
decay from \nuc{49}{Fe} to populate excited states in \nuc{49}{Mn}
\cite{Doss07}.

The measured inclusive one-proton pickup reactions cross sections on the
\nuc{12}{C} and \nuc{9}{Be} targets are $\sigma_{exp}^{inc}(\nuc{49}{Mn})
=2.00(13)$~mb and $\sigma_{exp}^{inc}(\nuc{49}{Mn})=1.63(8)$~mb, respectively.
These are comparable in magnitude to the earlier one-proton pickup cross
section to the $sd$-shell nucleus \nuc{21}{Na}, with a \nuc{9}{Be} target,
for which $\sigma_{exp}^{inc}=1.85(12)$ mb \cite{Gade07a}.

For the partial cross sections, by combining the single-particle cross
sections from the reaction model calculations (Section \ref{reaction})
with the associated GXPF1A shell-model spectroscopic factors, a distinctive
population pattern is predicted from the interplay of momentum matching
and proton shell structure. As can be seen from Table~\ref{tab:conf_49},
below the proton separation energy of \nuc{49}{Mn}, only the first
excited $7/2^-$ state is predicted to be populated with appreciable cross
section. Although the spectroscopic factors for pickup to the predicted
$1/2^-$ and $3/2^-$ states (proton $\ell=1$, $2p_{1/2}$ and $2p_{3/2}$
configurations) are comparable to $S_{SM}(f_{7/2})$, the single-particle
cross sections for pickup to the $\ell=1$ configurations are suppressed
by two orders of magnitude compared to those for the $\ell=3$ orbital,
independently of the target.

\begin{figure}[h]
\epsfxsize 8.5cm
\epsfbox{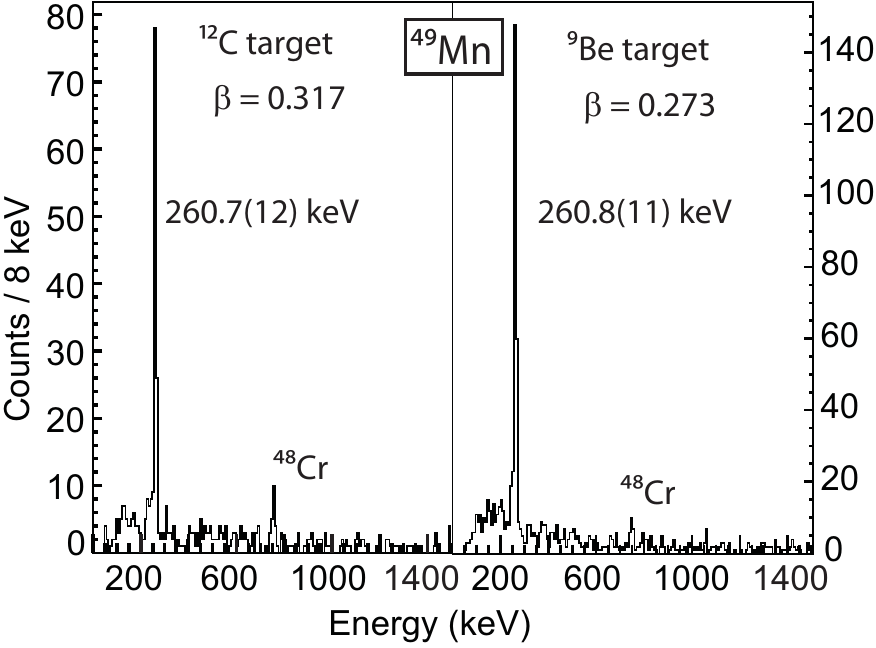}
\caption{\label{fig:Mn_gamma} Doppler-reconstructed $\gamma$-ray
spectra in coincidence with \nuc{49}{Mn} residues produced
in the one-proton pickup reactions of \nuc{48}{Cr} with \nuc{12}{C} (left)
and \nuc{9}{Be} (right) targets. For both targets, the only prominent
transition is the  decay of the first  excited $7/2^-$ state to the $5/2^-$
ground state. A peak, stemming from a tail of scattered \nuc{48}{Cr}
projectiles, is indicated. For the purpose of $\gamma$-ray spectroscopy, the
software gate was generous, to include all \nuc{49}{Mn} events, and therefore
included a small amount of \nuc{48}{Cr} scattered at large angles.}
\end{figure}

Figure~\ref{fig:Mn_gamma} (upper panel) shows the Doppler-reconstructed
in-beam $\gamma$-ray spectra detected in coincidence with \nuc{49}{Mn},
produced in one-proton pickup from the \nuc{12}{C} target (left) and
the \nuc{9}{Be} target (right). The only transition attributed to
\nuc{49}{Mn} is the de-excitation $\gamma$-ray from the first $7/2^-$
to the $5/2^-$ ground state, measured at $E=261$~keV. This confirms the
population pattern predicted by theory with almost all cross section
resulting from the pickup to the $\ell=3$, $7/2^-$ state at 262~keV
excitation energy. Within the level of the statistics, there is no
evidence for the decays of the $1/2^-$ and $3/2^-$ excited states
that are predicted to be only weakly populated in the reaction. From the
efficiency-corrected peak area relative to the number of \nuc{49}{Mn}
produced in the pickup, the partial cross section for the population
of the $7/2^-$ state could be extracted ($\sigma(7/2^-)=1.60(22)$~mb
for the \nuc{12}{C} target and $\sigma(7/2^-)=1.43(15)$~mb for the
\nuc{9}{Be} target). The partial cross section to the \nuc{49}{Mn}
ground state was derived by subtraction, i.e. $\sigma(g.s.)=
\sigma_{exp}^{inc}-\sigma(7/2^-)$. We note that the ground-state
cross section determined in this way is an upper limit since some
feeding from unobserved higher-lying proton-unbound states with a
$\gamma$-ray branch cannot be excluded. We obtain $\sigma(5/2^-_{gs})
\leq 0.40(25)$~mb for the \nuc{12}{C}-induced reaction and $\sigma
(5/2^-_{gs}) \leq 0.20(17)$~mb for the \nuc{9}{Be}-induced pickup.
Due to the significantly reduced $\gamma$-ray detection efficiency
at higher energies, we would not have been able to observe $\gamma$-ray
transitions from the predicted $1/2^-$ and $3/2^-$ states above
1700~keV, the strongest of the two being expected to be populated
at a level of less than 0.04~mb.

\begin{table}
\caption{Shell-model proton configurations $[\nuc{48}{Cr}(0^+) \otimes
n\ell_j]$ and their spectroscopic factors, $S_{SM}$, single-particle cross
sections to the individual final states in \nuc{49}{Mn} (summed over
the included \nuc{11}{B} and \nuc{8}{Li} final states), and the theoretical
cross sections $\sigma_{th}$ calculated using Eq.\ (\ref{eq:xsec}).}
\begin{ruledtabular}
\begin{tabular}{lccccccccc}
\multicolumn{6}{c}{ }& \multicolumn{2}{c}{\nuc{12}{C} target }&
\multicolumn{2}{c}{\nuc{9}{Be} target }\\
$J^{\pi}$ & $E_x$ &\multicolumn{3}{c}{ SM conf} &
$S_{SM}$ & $\sigma_{sp}$ & $\sigma_{th}$& $\sigma_{sp}$ & $\sigma_{th}$\\
& (keV) & & & & &  (mb) & (mb)&  (mb) & (mb)\\
\hline
$7/2^-$& 262 & &$[0^+ \otimes 1f_{7/2}]$ && 0.425 & 4.141 & 1.65 & 3.905 & 1.56\\
$1/2^-$& 1703 && $[0^+ \otimes 2p_{1/2}]$ && 0.245 & 0.040 & 0.009 & 0.039 & 0.009\\
$3/2^-$& 1741 & &$[0^+ \otimes 2p_{3/2}]$ && 0.373 & 0.114 & 0.04 &  0.082 & 0.03\\
\hline
\multicolumn{6}{l}{inclusive cross sections: {$\sigma_{th}^{inc}=\Sigma_j
  \sigma_{th}^j$ }} & & 1.70 &    &  1.60
\label{tab:conf_49}
\end{tabular}
\end{ruledtabular}
\end{table}

\begin{figure}[h]
\epsfxsize 8.0cm
\epsfbox{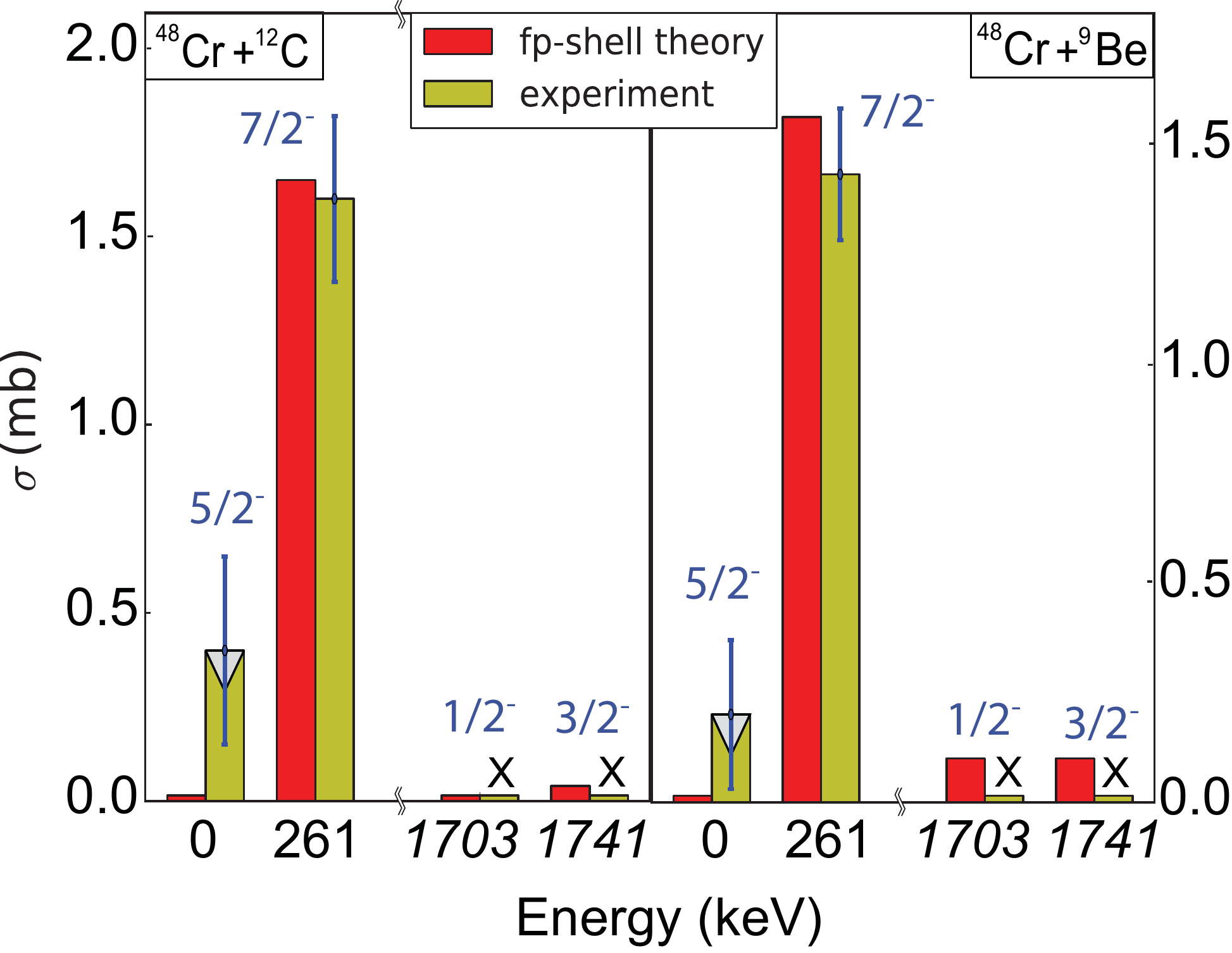}
\caption{\label{fig:Cr_Mn} (Color online) Comparison of measured and
calculated partial cross sections for proton pickup to \nuc{49}{Mn}.
The DWBA reaction model (see text) and the shell-model spectroscopic
factors from the GXPF1A effective interaction are used. The measured
partial cross section to the ground state represents an upper limit
obtained from the subtraction $\sigma(g.s) \leq \sigma_{exp}^{inc} -
\sigma(7/2^-)$. The expected higher-lying $2p$-shell states are
predicted to be populated only very weakly and could not be observed
in the present measurement (marked with x).}
\end{figure}

Figure~\ref{fig:Cr_Mn} compares the measured and calculated
partial cross sections for the \nuc{12}{C} and \nuc{9}{Be}-induced
one-proton pickup to \nuc{49}{Mn}. The measured data reproduces the
distinct population pattern with almost all reactions populating
the first $7/2^-$ excited state in \nuc{49}{Mn}. The measured and
calculated cross sections, $\sigma(7/2^-)$, agree very well within
the experimental uncertainty for both targets.

The measured inclusive cross section for the \nuc{12}{C} target,
$\sigma_{exp}^{inc}=2.00(13)$~mb, is about 18\% higher than the
calculated cross section. For the \nuc{9}{Be} target, the measured
and calculated inclusive cross sections agree, with $\sigma_{exp
}^{inc}=1.63(8)$~mb. The measurement also confirms the reaction
model prediction that the \nuc{12}{C}-induced pickup proceeds with
a slightly higher cross section.

\subsection{Proton pickup to \nuc{51}{Co}}
The most neutron-deficient even-$N$ Co isotope that is known to be
particle bound in its ground state, \nuc{51}{Co}, was first observed
in 1987~\cite{Pou87}. The ground state of this near-dripline nucleus
was tentatively assigned to be $7/2^-$~\cite{nndc}, with an evaluated
proton separation energy of only $S_p(\nuc{51}{Co})=90$~keV based
on systematics~\cite{AW2003}. This makes \nuc{51}{Co} the most
weakly-bound nucleus yet studied with a fast one-nucleon pickup
reaction. No bound or unbound excited states are reported in the
literature for~\nuc{51}{Co}. Based on the weak binding and the
non-observation of $\gamma$-ray transitions in coincidence with
\nuc{51}{Co}, we assume that, in the experiment, the cross section for
the population of the ($7/2^-$) ground state exhausts the inclusive
cross section. This is unlike the case of the one-proton pickup from
\nuc{22}{Mg} to \nuc{23}{Al}~\cite{Gade07b}, where a $\gamma$-ray decay
of an excited states above the proton separation energy was observed and
its contribution to the transfer cross section was included
in the calculations.

The measured inclusive cross sections on the \nuc{12}{C} and
\nuc{9}{Be} targets are $\sigma_{exp}^{inc}(\nuc{51}{Co})=0.53(13)$~mb
and $\sigma_{exp}^{inc}(\nuc{51}{Co})=0.57(8)$~mb, respectively. These
are comparable in magnitude to the proton pickup cross section to the
$sd$-shell nucleus \nuc{23}{Al}, measured on a \nuc{9}{Be} target, where
$\sigma_{exp}^{inc}=0.54(5)$~mb \cite{Gade07b}. We note that \nuc{23}{Al}
is also very weakly bound with $S_p(\nuc{23}{Al})=142.11(43)$~keV
\cite{Saa09}.

The weak binding of \nuc{51}{Co} makes this nucleus the ideal candidate
to study the one-proton pickup into the $f_{7/2}$ orbital. As outlined
above, we may assume, to a very good approximation,  that $\sigma_{exp
}^{inc}=\sigma(7/2^-)$. Table~\ref{tab:conf_51} summarizes the shell-model
spectroscopic strength for the assumed $7/2^-$ configuration and the
corresponding theoretical cross sections for the population of the
\nuc{51}{Co} ground state in one-proton pickup from the two targets.
In spite of the weak binding and rather small spectroscopic factor,
$S_{SM}=0.246$, the cross section is still predicted to be of the
order of 0.7~mb, which is sizable for the production of such an exotic,
weakly-bound nucleus.

\begin{table}
\caption{As Table~\ref{tab:conf_49} but for one-proton
pickup to \nuc{51}{Co}.}
\begin{ruledtabular}
\begin{tabular}{lcccccccccc}
\multicolumn{6}{c}{ }& \multicolumn{2}{c}{\nuc{12}{C} target }&
\multicolumn{2}{c}{\nuc{9}{Be} target }\\
$J^{\pi}$ & $E_x$ &\multicolumn{3}{c}{ SM conf} &
$S_{SM}$ & $\sigma_{sp}$ & $\sigma_{th}$& $\sigma_{sp}$ & $\sigma_{th}$\\\
  & (keV) & & & & &  (mb) & (mb) & (mb) & (mb)\\
\hline
$7/2^-$& 0 & &$[0^+ \otimes 1f_{7/2}]$ && 0.246 & 3.002 & 0.70& 2.858 & 0.66
\label{tab:conf_51}
\end{tabular}
\end{ruledtabular}
\end{table}

\begin{figure}[h]
\epsfxsize 7.2cm
\epsfbox{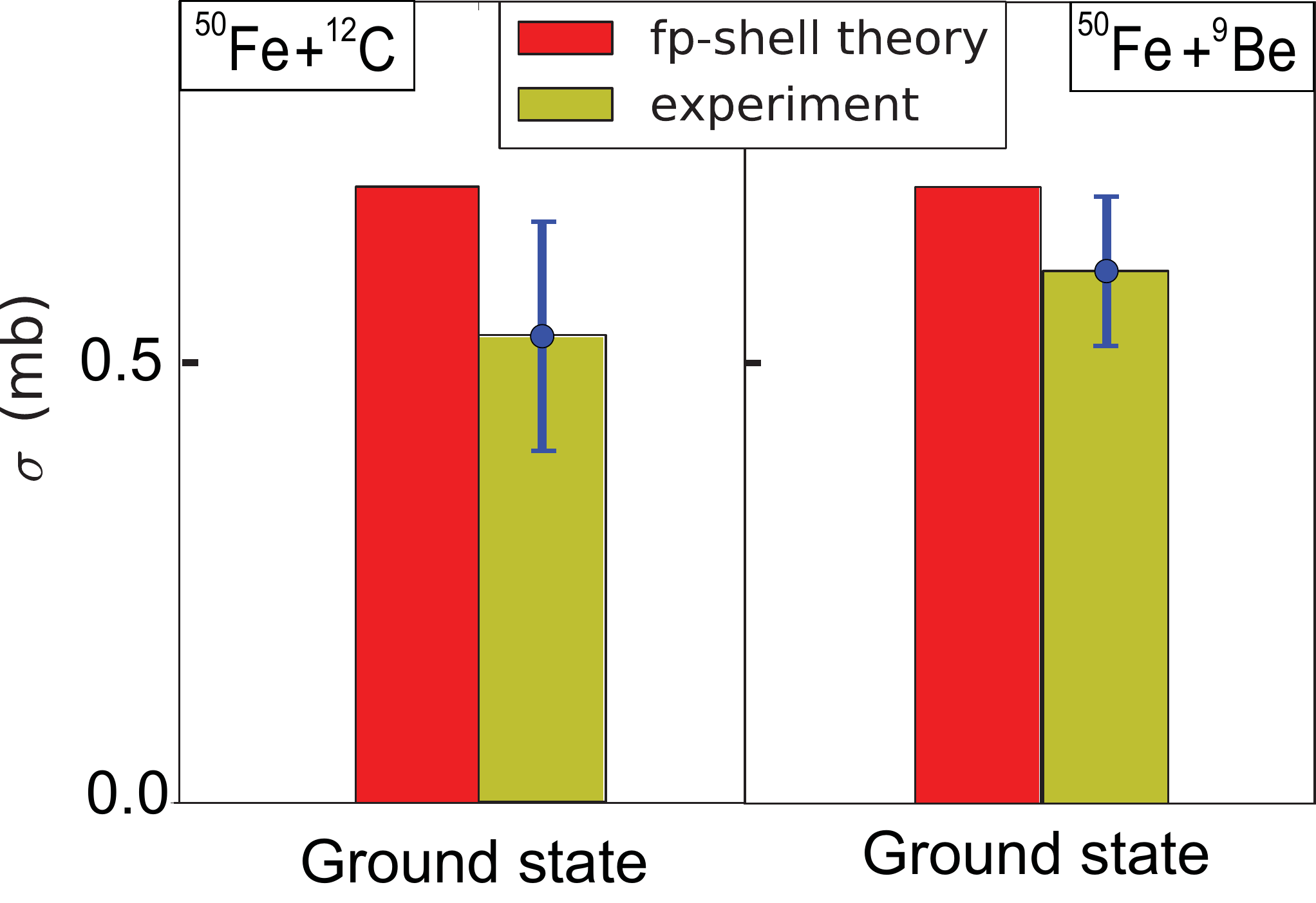}
\caption{\label{fig:Fe_Co} (Color online) Comparison of the measured
inclusive cross section and the calculated partial cross section for
one-proton pickup to  the ground state of \nuc{51}{Co}. The DWBA reaction
model (see text) and the shell-model spectroscopic factor from the GXPF1A
effective interaction are used. No transitions from excited states were
observed ($S_p=90$~keV).}
\end{figure}

Figure~\ref{fig:Fe_Co} compares the measured and calculated inclusive
cross sections for the $(7/2^-)$ ground state of \nuc{51}{Co}. Measurement
and the reaction model predictions agree within the experimental uncertainty
for the \nuc{9}{Be}-induced reaction while for the \nuc{12}{C}-induced
pickup the measurement is about 24\% lower than the prediction.

\subsection{Momentum distribution diagnostics}
In earlier work it has been pointed out that the longitudinal momentum
distributions of the projectile-like one-nucleon pickup residues can
be exploited as an indicator of contributions to the pickup reaction
mechanism that result in non-two-body final states \cite{Gade11}. This
diagnostic value arises since, for a strict two-body collision, the
reaction-induced momentum distribution of the residues arising from
kinematics and angular deflections of the reaction products is essentially
$\delta$-function-like compared to (a) the momentum spread of the incoming
beam of $\Delta p/p=1$\% and (b) the differential momentum loss of the
projectile and heavy residue in the thick target. The latter depends quite
sensitively on the interaction point in the reaction target. Thus, deviations
of the measured reaction residue momentum distributions from the momentum
profile of the projectile beam having passed through the target and folded
with the differential momentum loss of the mass ($A_p$+1)-residue -- a
rectangular function with a width that takes into account the difference in
momentum of the heavy residue depending on the pickup occurring at the front
or the back of the thick target -- will indicate non-two-body contributions.
This was demonstrated most clearly in Ref.\ \cite{Gade11}, for one-neutron
pickup reactions from a \nuc{9}{Be} target, populating excited, unbound
\nuc{8}{Be}$^*$ target residues.

\begin{figure}[h]
\epsfxsize 8.6cm
\epsfbox{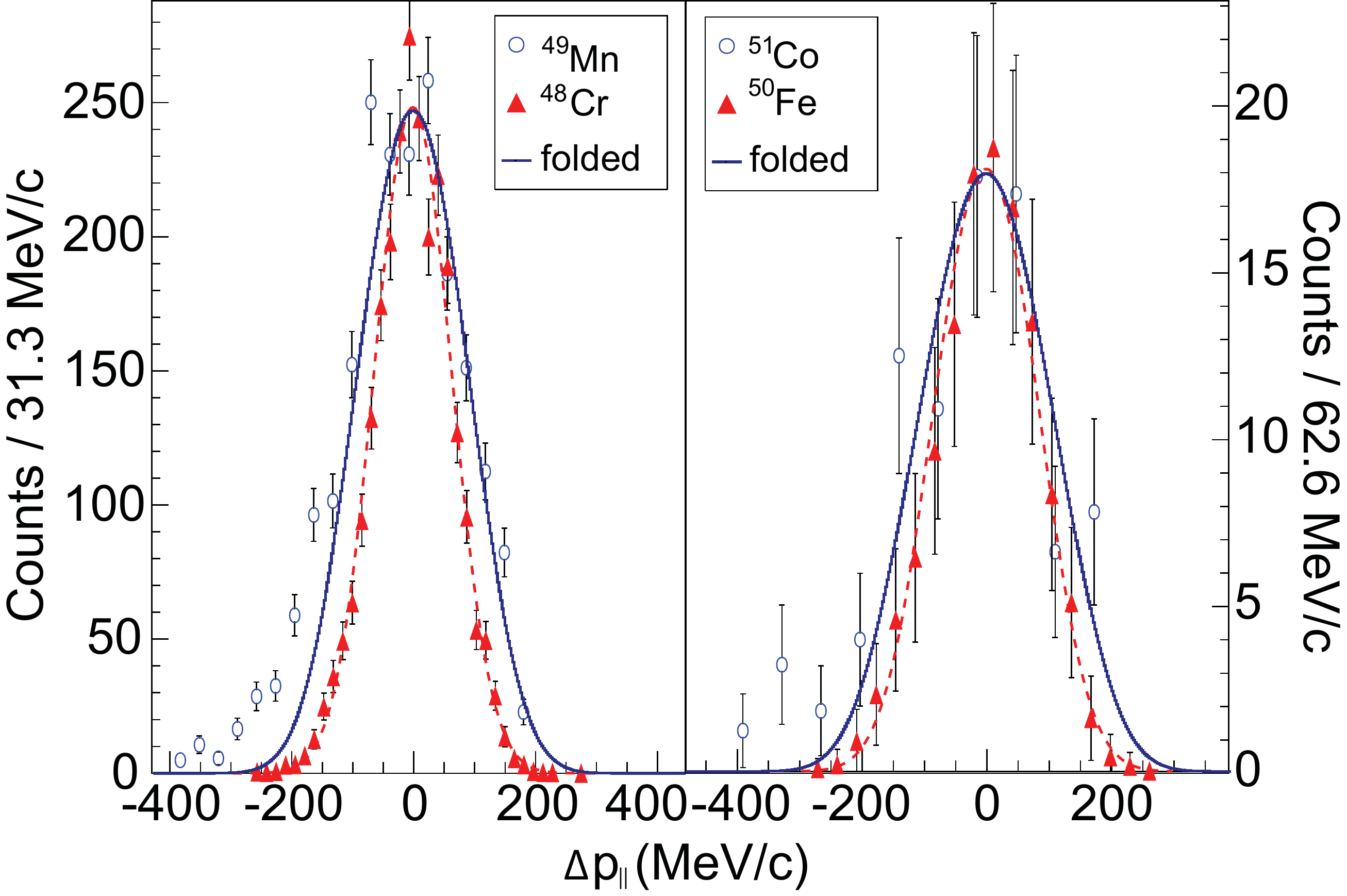}
\caption{\label{fig:MnCo_momentum} (Color online) Parallel momentum
distributions (blue symbols) of the \nuc{49}{Mn} and \nuc{51}{Co}
projectile-like residues following proton pickup on a \nuc{12}{C}
target. For comparison, the profiles of the unreacted \nuc{48}{Cr} and
\nuc{50}{Fe} projectiles are also shown (red symbols). The red dashed
line is a fit to these data points. The solid blue line is the result
of folding the projectile momentum profiles (red dashed lines) with a
rectangular distribution with a width that corresponds to the differential
momentum loss of the pickup reaction residues in the target, depending
on the interaction point. }
\end{figure}

Figure~\ref{fig:MnCo_momentum} shows the measured longitudinal momentum
distributions of the \nuc{49}{Mn} and \nuc{51}{Co} pickup reaction residues
produced on the \nuc{12}{C} target. Overlayed on the figure are also (i)
the momentum profiles of  the unreacted \nuc{48}{Cr} and \nuc{50}{Fe}
projectiles having passed through the target, and (ii) the result after
folding these profiles with the rectangle function that takes into account
the additional broadening of the distribution originating from the unknown
reaction point in the (thick) reaction target.

As can be seen, after folding the projectile momentum profiles with
the differential momentum loss experienced by the residues in the
target, the predicted parallel momentum distributions describe the
measured distributions very well. Small deviations are visible on the
low-momentum side of the \nuc{49}{Mn} momentum distribution where a
tail structure develops. This is in line with earlier one-proton
pickup reaction work on a \nuc{9}{Be} target \cite{Gade07a} and
one-neutron pickup reaction work on \nuc{12}{C} \cite{Gade11}, where
the momentum distributions show very little deviation from the strict
two-body reaction mechanism expectation.

\section{Summary and conclusion}
In summary, we have employed intermediate-energy one-proton pickup
reactions on \nuc{12}{C} and \nuc{9}{Be} targets to study nuclei in
the $fp$-shell. Our experimental approach combined thick reaction
targets, for luminosity, and $\gamma$-ray spectroscopy to tag the
final states in the reaction residues. Reaction model calculations
assuming one-step transfers and two-body channels were performed.
Proton pickups from \nuc{48}{Cr} to \nuc{49}{Mn} and from \nuc{50}{Fe}
to \nuc{51}{Co} were studied, each on both light targets. No bound
excited final states were observed in \nuc{51}{Co}. The measured
and calculated cross sections confirm that the \nuc{51}{Co} ground
state is almost certainly a $7/2^-$ state.

The measured cross sections were found to agree well with the two-body reaction
model calculations carried out in the distorted waves Born approximation.
Momentum-matching considerations, embedded in the reaction model dynamics,
predict a distinctive pattern of final states population in the heavy
residues, with a two-order of magnitude dominance of pickup into the
$\ell=3$, $1f_{7/2}$ proton orbital compared to pickup into the $\ell=1$,
$2p_j$ orbitals. The \nuc{49}{Mn} and \nuc{51}{Co} longitudinal momentum
distributions were also used to demonstrate that pickup events leading
to non-two-body final states could make only very small contributions
to the measured reaction cross sections.

The present study complements previous work, on the \nuc{9}{Be}-induced
one-proton pickup process and the \nuc{12}{C}-induced one-neutron pickup
reaction, and extends these studies into the $fp$-shell. The unique
selectivity of the reaction mechanism for the population of high-$\ell$
over low-$\ell$ orbitals was clearly shown, reinforcing expectations
of the potential of this reaction as a spectroscopic tool for the
identification of high-$\ell$ intruder states that often indicate the
breakdown of shell closures in nuclei far from stability.

\begin{acknowledgments}
This work was supported by the National Science Foundation under Grants
No. PHY-0606007 and PHY-0758099 and by the UK Science and Technology
Facilities Council under Grants ST/F012012/1 and ST/J000051/1. A.G. is
supported by the Alfred P. Sloan Foundation.
\end{acknowledgments}

\end{document}